\documentstyle[iopconf1,epsf]{article}

\newcommand{\be}[1]{\begin{equation} \label{(#1)}}
\newcommand{\ee}{\end{equation}}
\newcommand{\ba}[1]{\begin{eqnarray} \label{(#1)}}
\newcommand{\ea}{\end{eqnarray}}
\newcommand{\nn}{\nonumber}
\newcommand{\rf}[1]{(\ref{(#1)})}

\begin{document}
\title{New exotics in the double beta decay contributions zoo
\footnote[2]{Talk presented by Heinrich P\"as}}
\author{H.V. Klapdor--Kleingrothaus $^{1}$,
H. P\"as
\footnote[1]{E-mail:
Heinrich.Paes@mpi-hd.mpg.de}$^{1}$, 
and \\
U. Sarkar$^2$}
\affil{$^1$ Max--Planck--Institut f\"ur Kernphysik,\\ P.O. Box 103980,
D--69029 Heidelberg, Germany}
\affil{$^2$ Physical Research Laboratory, Ahmedabad, 380 009, India}

\beginabstract
We discuss the potential of neutrinoless double beta decay 
for testing Lorentz invariance and the weak
equivalence principle as well as contributions from dilaton exchange gravity
in the neutrino sector. While neutrino oscillation
bounds constrain the region of large mixing of the weak and
gravitational eigenstates, we obtain
new constraints on violations of Lorentz invariance and the equivalence 
principle from neutrinoless double beta decay, 
applying even in the case of no mixing. 
Double beta decay thus probes a totally unconstrained region in the parameter 
space. 
\endabstract

\section{Introduction}
Special relativity 
and the equivalence principle can be considered as the most 
basic foundations of the theory of gravity. 
However, string theories allow for or even predict 
the violation of these laws (see \cite{kost,nick} and references therein).
Many experiments hunt for these exotics (figure 1), which have been tested
to a very high 
level of
accuracy \cite{rel} for ordinary matter - generally for 
quarks and leptons of the first
generation. These precision tests of 
local Lorentz invariance -- violation of the equivalence 
principle should produce a similar effect \cite{will} -- probe for any 
dependence of the (non--gravitational) laws of physics on a laboratory's 
position, orientation or velocity relative to some preferred frame of
reference, such as the frame in which the cosmic microwave background is 
isotropic.  

A typical feature of the violation of local Lorentz invariance (VLI)
is that different species of matter have a characteristical 
maximum attainable velocity.
This can be tested in various sectors of the standard model
through vacuum Cerenkov radiation \cite{gasp}, photon decay \cite{cole},
neutrino oscillations \cite{glash,nu1,nu2,hal,nu3,lisi} and $K-$physics
\cite{hambye,vepk}. In this article we extend
these arguments to derive new constraints from neutrinoless double
beta decay. 

\begin{figure}
\epsfysize=80mm
\vspace*{5mm}
\hspace*{1cm}
\epsfbox{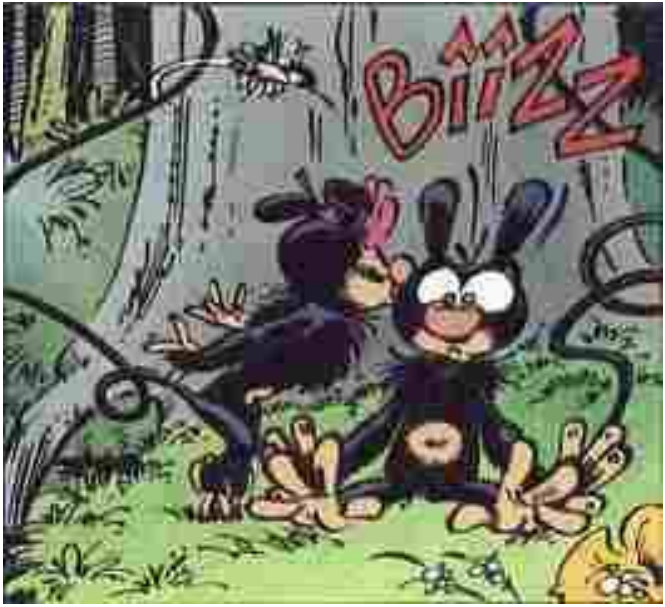}
\vspace*{5mm}
\caption{\it 
The hunt for exotics 
is still going on and may be successful in neutrinoless double beta decay
\label{fig1}}
\end{figure}

The equivalence principle implies that spacetime is described by
unique operational geometry and hence universality of the gravitational 
coupling for all species of matter. In the recent years there
have been attempts to constrain a possible amount of 
violation of the equivalence principle (VEP) in the neutrino sector
from neutrino oscillation experiments \cite{nu1,nu2,hal,nu3,lisi}.
However, these bounds don't apply when the gravitational and the
weak eigenstates have small mixing. In the following we point out
that neutrinoless double beta decay also constrains  
VEP. VEP implies different neutrino species to feel  
different gravitational potentials while propagating through the 
nucleus and hence the contributions of different eigenvalues don't cancel
for the same effective momentum. 
The main result is that neutrinoless double beta decay can constrain
the amount of VEP even if the mixing angle is zero, i.e.
if only the weak equivalence principle is violated, for which 
no bound exists at present.

\section{Violations of Lorentz invariance}
For sake of clarity we formulate the problem for a
two generation scenario involving $\nu_e$ and $\nu_x$ with 
$x=\mu,\tau,s$.
Neutrinos of different species may have different maximum attainable 
velocities if there is violation of local Lorentz invariance
(VLI) and hence violation of special relativity \cite{cole}.
We first assume that the weak eigenstates cannot be diagonalized 
simultaneously with the velocity eigenstates and the neutrinos are
relativistic point particles. 
The effective Hamiltonian in the weak basis $[\nu_e ~~ \nu_x ]$ is
\begin{equation}
H = U_{m} H_{m} U_{m}^{-1} + U_v H_v U_v^{-1} \label{h}.
\end{equation}
In absence of VLI the neutrino mass matrix
in the mass basis $[\nu_1 ~~ \nu_2 ]$ is given by
\be{hsew}
H_{m} = \frac{(M_{m})^2}{2 p} = \frac{1}{2 p} {\pmatrix{
m_1 & 0 \cr 0 & m_2 }}^2 \label{hsew}
\end{equation}
and the VLI part of the hamiltonian as
\begin{equation}
H_v = \pmatrix{ v_1 & 0 \cr 0 & v_2} p  , 
\ee
to leading order in $\bar{m}^2/p^2$. Here p denotes the momentum and
$\bar{m}$ the average mass, and for any quantity $X$ we define $\delta X
\equiv(X_1-X_2)$, $\bar{X} = (X_1+X_2)/2$. 

In the absence of VLI, i.e. if the special theory of
relativity is valid, $v_i = 1$, and $H_v$ simply becomes the momentum
of the neutrinos. Here we are interested in a single neutrino
beam (for neutrino oscillation experiments) or a single virtual neutrino
propagating inside the nucleus with a particular momentum. For this 
reason we assume the momenta of both the neutrinos are $p$. Then
$v_i $ corresponds to the maximum attainable velocity of the corresponding
momentum eigenstates. Hence $v_1 - v_2 = \delta v$ is a measure of 
VLI in the neutrino sector. As typical or ``standard'' maximum attainable 
velocity $\frac{v_1+v_2}{2}=1$ is assumed. All previous bounds on this quantity
$\delta v$ in the neutrino sector were derived from neutrino oscillation
experiments and for that reason these bounds are valid only for 
large gravitational mixing. As we shall point out, 
neutrinoless double beta decay can constrain
$\delta v$ even if the mixing angle vanishes. 

We shall not consider any $CP$ violation, and hence $H_{m}$ and
$H_v$ are real symmetric matrices and $U_{m}$ and $U_v$ are orthogonal
matrices $U^{-1}= U^T$. They can be parametrized as 
$U_i = \pmatrix{\cos \theta_i & \sin \theta_i \cr -\sin \theta_i
& \cos \theta_i}$, where $\theta_i$ represents weak mixing angle $\theta_m$ 
or velocity mixing angle $\theta_v$. We can now write down the 
weak Hamiltonian $H_w$ in the basis $[\nu_e ~~ \nu_x]$, 
in which the charged lepton mass matrix 
and the charged current interaction are diagonal:
$$
H = p I +  \frac{1}{2 p} { \pmatrix{ M_{+} & M_{12} \cr
M_{12} & M_{-} }}^2.
$$
Here $I$ is the identity matrix and
\ba{obs}
M_{+} &=& \bar{m} \pm \frac{cos2\theta_m}{2}
\delta m \nn \\
&&\pm \frac{p^2}{\bar{m}} \delta v \left( \frac
{\cos2\theta_v}{2} - {\delta m \over 4 \bar{m}} \cos 2 (\theta_m - \theta_v)
\right) 
\ea
is the double beta observable in the presence of VLI.
In the mass mechanism of neutrinoless double beta decay, 
the half life
\be{t12}
[T_{1/2}^{0\nu\beta\beta}]^{-1}=\frac{\langle m \rangle^2}{m_e^2} 
G_{01} |ME|^2
\ee
is proportional to the effective neutrino mass 
$\langle m \rangle=m_{ee}=M_+$.
Here $ME$ denotes the nuclear matrix element $ME=M_F-M_{GT}$, $G_{01}$
corresponds to the phase space factor defined in \cite{doi} and $m_e$ is the 
electron mass.
The double beta observable can be written as
\ba{1}
<m> &=& \sum_i U_{e i}^2 m_i = 
m_1 \cos^2 \theta_w + m_2 \sin^2 \theta_w\nn \\
&=& \bar{m} + {1 \over 2} \delta m \cos 2 \theta_w.
\ea
If $m_{ee}= 0$, the two physical eigenstates with
eigenvalues $m_1$ and $m_2$ will contribute to the neutrinoless 
double beta decay by an amount $U_{e1}^2 m_1$ and $U_{e2}^2 m_2$,
respectively, which cancels each other. However, if these two physical 
states have different maximum attainable velocities, corresponding to VLI,
this cancellation 
will not be exact for the same cut--off effective momentum in the neutrino 
propagator. As a result, even if 
$m_{ee}= 0$, a non-vanishing contribution to
neutrinoless double beta decay can exist, which is
proportional to the amount of VLI. In this case the double beta observable is
given by $M_+$ in \rf{obs}. From \rf{obs} it can easily be seen that in the 
region of maximal mixing,
$\cos 2 \theta_v = 0$, the double beta decay rate vanishes. 
Thus neutrinoless double beta decay doesn't constrain
the amount of VLI for maximal mixing. However, if the mixing approaches zero,
the most stringent bound from neutrinoless double 
beta decay is obtained. 
In this case $\delta{v}/2$ can be understood as deviation from the standard
maximum attainable velocity $\bar{v}$.
Obviously, for the case of vanishing mixing 
neutrino oscillation experiments cannot give any bound on the amount of 
VLI: Allowing only for VLI without mixing will not imply neutrino 
oscillations. 

To give a bound on VLI in the small mixing region 
(including $\theta_v=\theta_m\simeq 0$)
we assume conservatively
$\langle m \rangle \simeq 0$.
We also assume 
$\delta{m} \leq \bar{m}$, and thus $\frac{\delta m}{4 \bar{m}}$ may be 
neglected. 
Due to the $p^2$ enhancement the nuclear matrix element of the 
mass mechanism have to be replaced by $\frac{m_p}{R}\cdot 
(M_F^{'}-M_{GT}^{'})$ with the nuclear radius $R$ and the proton mass $m_p$, 
which has been 
calculated in \cite{mat}.
Inserting the recent conservative
half life limit obtained from the Heidelberg--Moscow 
experiment \cite{double}, $T_{1/2}^{0\nu\beta\beta}>1.8\cdot 10^{25} y
~~(90 \% C.L.)$,  
a bound on the amount of VLI as a function of the average neutrino mass
$\bar{m}$ can be given. The most reliable assumption for $\bar{m}$ is obtained 
from the cosmological bound $\sum_i m_i < 40$ eV \cite{raf}, i.e., 
$\bar{m}<13$ eV for three generations, implying a bound of \cite{beta}
$$\delta v < 4 \times 10^{-16}~~~~ {\rm for}~~~ \theta_v=\theta_m \simeq0.$$ 

In figure 2 the bound implied by double beta decay
is presented for the entire range of $sin^2 2 \theta_v$ and compared with
bounds obtained from neutrino oscillation experiments in \cite{hal}. 
It should be stressed also that the GENIUS proposal of the Heidelberg group
\cite{gen} could improve these bounds  
by about 1--2 orders of magnitude \cite{beta}.

\begin{figure}[!t]
\epsfysize=70mm
\hspace*{1cm}
\epsfbox{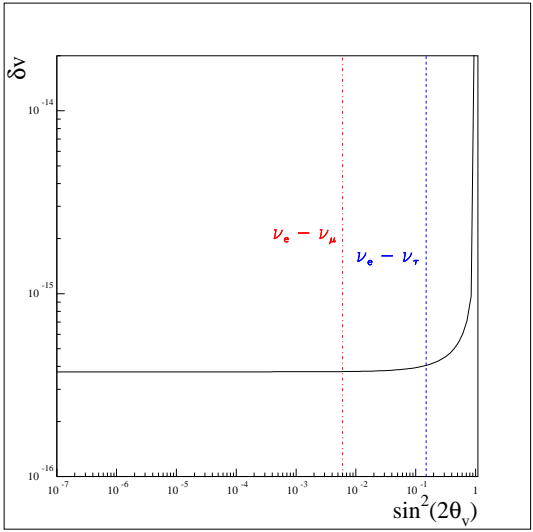}
\caption{\it Double beta decay bound (solid line)
on violation of Lorentz invariance 
in the neutrino sector, excluding the region to the upper left. 
Shown is a double logarithmic plot 
in the $\delta v$--$\sin^2(2 \theta)$ parameter space. 
The bound becomes most stringent for the
small mixing region, which has not been constrained from any
other experiments. For comparison the bounds obtained from neutrino oscillation
experiments (from [10])
in the $\nu_{e} - \nu_{\tau}$ (dashed lines) and in the
$\nu_e - \nu_\mu$ (dashed-dotted lines) channel, excluding the region to the 
right, are shown. 
\label{fig2}}
\vspace*{-5mm}
\end{figure}

\section{Violations of the equivalence principle}
 
In the following we present the formalism for violation of the equivalence 
principle (VEP). While in the final expression the amount of
VLI just will be replaced by the corresponding term of
VEP, the effects may have a totally different 
origin.  In a linearized 
theory the
gravitational part of the Lagrangian to first order in a 
weak gravitational field $g_{\mu\nu}=\eta_{\mu\nu}+    h_{\mu\nu}$
($h_{\mu\nu}= 2\frac{\phi}{c^2}  {\mbox diag}(1,1,1,1)$)
can be expressed as ${\cal L} = -\frac{1}{2}(1+g_i)h_{\mu\nu}T^{\mu\nu}$,
where  $T^{\mu\nu}$  is the  stress-energy  in the  gravitational
eigenbasis. In presence of VEP the couplings $g_i$ may differ. Assuming only 
violation of the weak
equivalence principle, the gravitational
interaction is diagonal. In this case there 
does not exist any bound on the amount of VEP in the neutrino sector. 
We point out that this
region of the parameter space is most restrictively bounded by  
neutrinoless double beta decay.

The effective Hamiltonian in the weak basis again can be written as
\begin{equation}
H = p  I + U_m H_{m} U_m^{-1} + U_G H_G U_G^{-1}, \label{hha}
\end{equation}
with $H_m$ given in \rf{hsew} and
\ba{4}
H_G &=& \pmatrix{ G_1 & 0 \cr 0 & G_2} \nn \\
&=& \pmatrix{
- 2 (1 + g_1) \phi (p + \frac{\bar{m}^2}{2 p}) & 0 \cr 0 &
- 2 (1 + g_2) \phi (p + \frac{\bar{m}^2}{2 p})} \label{hga}
\ea
to first order in $\bar{m}^2/p^2$.
The resulting double beta decay observable is
\ba{5}
M_{+} &=& \bar{m} \pm {\cos 2 \theta_m \over 2} {\delta {m}} \nn \\
&&
\pm \frac{p}{\bar{m}} \delta G \left( \frac{\cos 2\theta_G}{2}
- {\delta m \over 4 \bar{m}} \cos (\theta_m - \theta_v) \right).
\ea

Compared to VLI, the expression remains unchanged for the case of VEP, 
except for replacing $\delta v$ by $\frac{1}{p}\delta G$.
Again, $\bar{g}=\frac{g_1+g_2}{2}$ can be considered as the standard 
gravitational coupling, for which the equivalence principle applies.

Thus the discussion of VLI can be directly translated to 
the VEP case and the bound from neutrinoless double beta
decay for $\theta_v=\theta_m =0$ is now given by \cite{beta}
\ba{99}
\phi \delta g < 4 \times 10^{-16} ~ ({\rm for~} \bar{m}<13 
{\rm eV})
\ea
In this case, $\delta G = p \phi \delta g$, where $\phi$ is the 
background Newtonian gravitational potential on the surface of
the earth. A natural choice for $\phi$ would be the earth's 
gravitational potential $(\sim 10^{-9})$, but another well motivated 
choice could be 
the potential due to all forms of distant matter. Unlike the case of
VLI, the bound on the VEP will depend on what one chooses for the
Newtonian potential $\phi$. For this reason, here we only present
the combined bound on $\phi \delta g$.

\section{Dilaton exchange gravity}

Recently it has been argued by Damour and Polyakov \cite{pd}
that string theory may lead to a new scalar type 
gravitational interaction via couplings of the dilaton
field and subsequently its consequence to neutrino 
oscillation has been studied \cite{dil}.
Damour and Polyakov have shown that the massless dilaton 
interaction modifies the gravitational potential energy and
there is an additional contribution from spin-0 exchange,
which results in a scalar type gravitational interaction \cite{pd}. 
The resulting theory is of scalar-tensor type with the two particle
static gravitational energy
\begin{equation} 
V(r)=- G_N m_A m_B (1+\alpha _A \alpha_B)/r.
\end{equation} 
Here $G_N$ is Newton's gravitational constant and $\alpha_j$ denotes the
couplings of the dilaton field $\phi$ 
to the matter field $\psi_j$, leading to a gravitational energy of
\begin{equation} 
L= m_j \alpha_j \overline{\psi_j}\psi_j \phi.
\end{equation} 
Thus the modified effective mass matrix of the neutrinos is now
given by \cite{dil}
\begin{equation} 
m^{(*)}=m-m \alpha \phi_c
\end{equation} 
where the classical value of the dilaton field $\phi_c=\phi_N 
\alpha_{ext}$ is characterized by the $\alpha$ value of the bulk 
matter producing it and a static matter distribution 
proportional to the Newtonian potential $\phi_N$ is assumed. 

The effective mass squared difference 
\begin{equation} 
\Delta m^{(*)2}=-2 \bar{m}^2 \phi_N \alpha_{ext} \delta \alpha
\end{equation} 
(for almost degenerate masses $m_1 \sim m_2 \sim \bar{m}$)
gives rise to neutrino oscillation.
The corresponding effect for $0\nu\beta\beta$ decay is
obtained by replacing $\delta g^S = 2 \delta \alpha^S \bar{m}^2$.
Comparing the arguments in the oscillations propabilities we get
\cite{dilbeta}
\begin{equation} 
M_+=\bar{m} \alpha_{ext} \Phi_N \delta \alpha 
\frac{\cos(2 \theta_G)}{2}.
\end{equation} 

In this case it is difficult to obtain any bound from neutrino
experiments, since for $\alpha_{ext}$ only upper bounds exist. 
To get an idea of the constraints obtainable from
neutrino experiments, if in the future $\alpha_{ext}$ is known, 
according to ref. \cite{dil} we assume $\phi_N=3 \cdot 10^{-5}$,
$\alpha_{ext}=\sqrt{10^{-3}}$ and $\bar{m}=2.5$ eV (as an upper 
bound obtained from tritium beta decay experiments \cite{tritium}).
In this case the quantity $\delta \alpha$ can not be constrained
from neutrinoless double beta decay.

\section{A comment on Halprin's critics}

In ref \cite{wrong} it is claimed that neither violations of
Lorentz invariance nor violations of the equivalence principle may
give sizable contributions to neutrinoless double beta decay. 
The argument discussed is the following: One considers the neutrino
propagator 
\begin{equation} 
\int d^4 q \frac{e^{-i q (x-y)} \langle m \rangle c_a^2}{m^2 c_a^4 
- q_0^2 c_a^2 + \vec{q}^2 c_a^2 }
\end{equation} 
with the standard $0\nu\beta\beta$ observable $\langle m \rangle$, the
neutrino four momentum $q$ and the characteristic maximal velocity $c_a$.
If one would neglect now $q_0$ and $m$ in the denominator, $c_a$ drops
out and the decay rate is independent of $c_a$.
The discussion of VEP makes use of the same argument.

However, in \cite{beta} it has been shown starting from the Hamiltonian level
that the propagator (or the $0\nu\beta\beta$ observable) is changed 
itself if one allows for Lorentz invariance violation. Since
\begin{eqnarray} 
H&=&\vec{q} c_a + \frac{m^2 c_a^4}{2 \vec{q} c_a} \nonumber  \\
&=& \vec{q} I + \frac {m^{(*)2} c_a^4}{2 \vec{q} c_a} \label{1}
\end{eqnarray} 
with $c_a = I + \delta v$ and $m^{(*)2}=m^2 + 2 \vec{q}^2 c_a \delta v $
an additional contribution to the effective mass
is obtained $\propto \vec{q}^2 \delta v$.
This mass-like term has a $\vec{q^2}$ enhancement and is not
proportional to the small neutrino mass.
This consideration answers also the frequently asked question ``What
is the source of lepton number violation?'' in this mechanism.
Comparable to a usual mass term, which can be both of Majoran type as well
as Dirac type the mass-like term 
$2 \vec{q}^2 c_a\delta v $ can be of Majorana type and act as the source
of lepton number violation in this context.

\section{Summary}

We discussed the potential of neutrinoless double beta decay searches for 
exotic phenomena such as violations of Lorentz invariance, the equivalence 
principle and contributions of dilaton exchange gravity.
We pointed out that 
neutrinoless double beta decay can constrain the amount of VLI or 
VEP. In particular, when the mixing of the gravitational 
eigenstates vanishes, the bounds from
neutrinoless double beta decay become most stringent, while this region is not 
constrained by any other experiments.

\section*{Acknowledgement}
We thank Gautam Bhattacharyya for useful discussions.

\end{document}